# Real-Time Patient-Specific ECG Classification by 1D Self-Operational Neural Networks

Junaid Malik, Ozer Can Devecioglu, Serkan Kiranyaz *Senior, IEEE*, Turker Ince, and Moncef Gabbouj, *Fellow*, *IEEE*.

*Abstract*— Despite the proliferation of numerous deep learning methods proposed for generic ECG classification and arrhythmia detection, compact systems with the real-time ability and high accuracy for classifying patient-specific ECG are still few. Particularly, the scarcity of patient-specific data poses an ultimate challenge to any classifier. Recently, compact 1D Convolutional Neural Networks (CNNs) have achieved the *state-of-the-art* performance level for the accurate classification of ventricular and supraventricular ectopic beats. However, several studies have demonstrated the fact that the learning performance of the conventional CNNs is limited because they are homogenous networks with a basic (linear) neuron model. In order to address this deficiency and further boost the patient-specific ECG classification performance, in this study, we propose 1D Self-organized Operational Neural Networks (1D Self-ONNs). Due to its self-organization capability, Self-ONNs have the utmost advantage and superiority over conventional ONNs where the prior operator search within the operator set library to find the best possible set of operators is entirely avoided. As the first study where 1D Self-ONNs are ever proposed for a classification task, our results over the MIT-BIH arrhythmia benchmark database demonstrate that 1D Self-ONNs can surpass 1D CNNs with a significant margin while having a similar computational complexity. Under AAMI recommendations and with minimal common training data used, over the entire MIT-BIH dataset 1D Self-ONNs have achieved 98% and 99.04% average accuracies, 76.6% and 93.7% average F1 scores on supra-ventricular and ventricular ectopic beat (VEB) classifications, respectively, which is the highest performance level ever reported.

## I. INTRODUCTION

Cardiac arrhythmia as being the most common cardiovascular disease poses the leading cause of mortality in the World [1]-[3]. In ECG, the sequence of heartbeats in each cardiac cycle exhibits individual electrical depolarization-repolarization patterns of the heart. The presence of an arrhythmia can be detected by an expert cardiologist by assessing a recorded or acquired ECG signal as the anomaly over the heart rate or rhythm or change in the morphological pattern. This is usually a tedious, subjective, and labor-intensive task. Numerous studies proposed several methods for automatic and accurate detection of arrhythmia over ECG signals. Earlier works [5]-[13] based on traditional signal processing and machine learning methodologies have not performed well for clinical use. The main reason behind this is that among different patients or even for the same patient but under different temporal, psychological, and physical conditions, significant variations may occur in the morphological characteristics and temporal/structural dynamics of ECG signals. During different times and under different circumstances, even the shape of each ECG beat, the QRS complex, P waves, and R-R intervals of a healthy individual will not be the same [4]. Hence such hand-crafted feature extraction may fail to capture the characteristics of each ECG beat variation for accurate classification. This is one of the reasons for their unreliable performance level for clinical usage, and their performance level varies significantly in large ECG datasets, [14], [15]. Other reasons can be the variation in severity of noise, the usage of different ECG sensors, inter-patient ECG signal variations, and differences in the prevalence of arrhythmia between databases.

Numerous deep learning-based methods have recently been proposed for ECG classification [16]-[22]. These methods are based on deep CNNs with high complexity and require massive labeled ECG data for training (e.g. > 50K beats). Moreover, as they require special parallelized hardware for proper functioning, they are not directly implementable on low-power or mobile devices. Furthermore, such methods are generic (one classifier for all patients) and thus are not immune from the aforementioned intra- and inter-patient variations. Moreover, such deep networks cannot be trained for a single patient due to the scarcity of the labeled data. Another major problem is the lack of common practice when a particular method is evaluated over a benchmark dataset since most of these methods vary the choice of the train and test data. To address this need, the *Association for the Advancement of Medical Instrumentation* (AAMI) recommends certain standards for evaluating the performance of the arrhythmia detection methods [23]. However, among the numerous methods proposed in the literature, only a few [13], [24]-[38] have followed the AAMI standards in their studies and even fewer have tested over the complete data from the benchmark MIT-BIH arrhythmia database [59]. This is indeed a crucial point for a fair and standardized comparison.

To address the aforementioned issues, in this study, we draw the focus on patient-specific ECG classification where the objective is to maximize the arrhythmia detection performance when the data is scarce and the network complexity is minimized for a real-time

S. Kiranyaz is with Electrical Engineering, College of Engineering, Qatar University, Qatar; e-mail: mkiranyaz@qu.edu.qa.

T. Ince are with the Electrical & Electronics Engineering Department, Izmir University of Economics, Turkey; e-mails: turker.ince@izmirekonomi.edu.tr.

J. Malik, O. C. Devecioglu and M. Gabbouj are with the Department of Computing Sciences, Tampere University, Finland; e-mails: junaid.malik@tuni.fi, ozer.devecioglu@tuni.fi and Moncef.gabbouj@tuni.fi.

application over any platform. Among many patient-specific ECG classification systems, [13], [15], [25]-[38], the landmark study in [27] proposed for the first time, a compact 1D CNN for real-time ECG classification and achieved the *state-of-the-art* performance demonstrated in MIT-BIH dataset while following the AAMI recommendations, i.e., for the training of each "patient-specific" classifier, only the first 5-min section from the beginning of each patient record together with the 245 beats randomly selected from the train partition of the MIT-BIH dataset should be used. As stated in [27], a dedicated 1D CNN can be easily trained for each patient and as a compact classifier, it can perform arrhythmia detection and classification task with utmost speed (requiring only a few hundreds of 1D convolutions). As a result, 1D CNNs were indeed the best choice especially for real-time advance warning and ECG monitoring on lightweight devices.

However, recent studies [40]-[46] have pointed out the fact that CNNs, similar to their predecessors, the Multi-Layer Perceptrons (MLPs), are homogenous networks with the sole linear neuron model from the 1950s (McCulloch-Pitts) [39]. This ancient neuron model is a crude model of the biological neurons or mammalian neural systems, which are highly heterogeneous and consist of diverse neuron types with specialized electrophysiological and biochemical properties [50]-[55]. Accordingly, MLPs and their popular derivatives, CNNs having such a homogenous network configuration based on such crude neuron model are capable of learning for relatively simple and linearly separable problems; however, they entirely fail to do so whenever the solution space of the problem is highly nonlinear and complex [40]-[46]. Thereafter Operational Neural Networks (ONNs) [46] have been proposed and like their predecessor Generalized Operational Perceptrons (GOPs) [40]-[45], they are heterogeneous networks with a non-linear neuron model which gives them an elegant diversity level to learn highly complex and multi-modal functions or spaces with minimal network complexity and training data. Recent studies [56], [57] have proposed the latest ONN variants, 2D Self-ONNs[1] for various image processing and regression tasks, and demonstrated that 2D Self-ONNs even with less number of neurons can achieve a superior learning performance whilst the performance gap between ONNs and CNNs widens further.

In this study, we propose 1D Self-organized ONNs (Self-ONNs) with the *generative* neuron model for patient-specific ECG classification. Hence, our objective is to achieve a superior ECG beat classification performance compared to the compact 1D CNNs [27] while keeping a similar network complexity. As the first study where 1D Self-ONNs have been proposed for a classification task, we aim to demonstrate the Self-ONNs potential on ECG classification and arrhythmia detection. 1D Self-ONNs have crucial advantages over conventional CNNs and ONNs. As illustrated in Figure 1, the convolutional and operational neurons of a CNN and an ONN have *fixed* nodal operators (linear and sinusoidal, respectively) in their 1x3 kernels. For 1D Self-ONNs *the generated* nodal function, **Ψ**, during training for each kernel element can be any arbitrary function. So, Self-ONNs neither need an operator set library in advance, nor require any prior search process to find the optimal nodal operator. We aim to demonstrate that this indeed yields a superior operational diversity and flexibility and in turn, a higher classification performance can be achieved using a very compact network model.

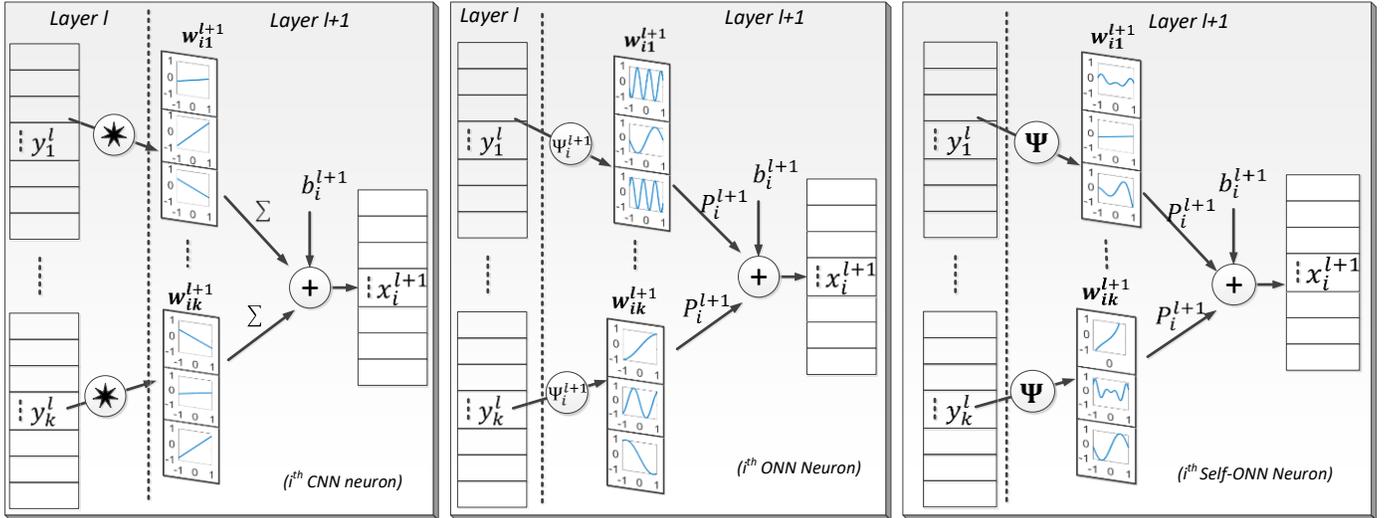

**Figure 1: An illustration of the 1D nodal operations with the 1D kernels of the $k^{th}$ CNN (left), ONN (middle), and Self-ONN (right) neurons at layer $l$.**

The rest of the paper is organized as follows: Section II presents 1D Self-ONNs with generative neurons in detail, formulates the forward-propagation (FP), the back-propagation (BP) training, their computational complexity and introduces the MIT-BIH dataset. Section III presents the experimental setup used for testing and evaluation of the proposed patient-specific ECG classification approach based on 1D Self-ONNs. Then, the overall results obtained from the ECG classification experiments are presented and comparative evaluations using the standard performance metrics against several state-of-the-art techniques are provided. Section IV makes a detailed discussion on the results and comparative evaluations. Finally, Section V concludes the paper and suggests topics for future research. We will briefly present the conventional (2D) ONNs to clarify the background and the terminology in Appendix A. In Appendix B, we will present the distribution of the beat

---
[1] The optimized PyTorch implementation of Self-ONNs is publically shared in http://selfonn.net/ .

classes and the detailed classification results per patient using the standard performance metrics.

## II. METHODS AND DATASET

In this section, we will cover the mathematical model of the proposed 1D Self-ONNs will be presented. To conclude, a simplification of the generative neuron will be discussed which can significantly reduce the computational cost by enabling the use of fast vectorized operations.

### A. 1D Self-organized Operational Neural Networks

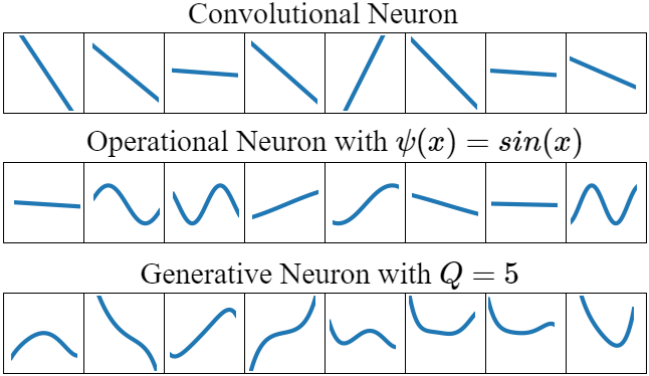

**Figure 2:** A visual comparison of different nodal transformation profiles entailed by the kernel of a convolutional, operational, and the proposed generative neuron of order Q. The generative neuron model enables enhanced nonlinearity and heterogeneity within the kernels.

Let us start with considering the case of the $k^{th}$ neuron in the $l^{th}$ layer of a 1D CNN. For the sake of brevity, we assume the same convolution operation with unit stride and the required amount of zero paddings. The output of this neuron can be formulated as follows:

$$x_k^l = b_k^l + \sum_{i=0}^{N_{l-1}} x_{ik}^l \quad (1)$$

where $b_k^l$ is the bias associated with this neuron and $x_{ik}^l$ is defined as:

$$x_{ik}^l = Conv1D(w_{ik}, y_i^{l-1}) \quad (2)$$

Here, $w_{ik} \in \mathbb{R}^K$ is the kernel connecting the $i^{th}$ neuron of $(l-1)^{th}$ layer to the $k^{th}$ neuron of the $l^{th}$ layer, while $x_{ik}^l \in \mathbb{R}^M$ is the input map, and $y_i^{l-1} \in \mathbb{R}^M$ are the $l^{th}$ and $(l-1)^{th}$ layers' $k^{th}$ and $i^{th}$ neurons' outputs, respectively. By definition, the convolution operation of (2) can be expressed as,

$$x_{ik}^l(m) = \sum_{r=0}^{K-1} w_{ik}^l(r) y_i^{l-1}(m+r) \quad (3)$$

The core idea behind an operational neuron is a generalization of the above as follows:

$$\overline{x_{ik}^l}(m) = P_k^l \left( \psi_k^l \left( w_{ik}^l(r), y_i^{l-1}(m+r) \right) \right)_{r=0}^{K-1} \quad (4)$$

where $\psi_l^k(\cdot) : \mathbb{R}^{M \times K} \to \mathbb{R}^{M \times K}$ and $P_k^l(\cdot) : \mathbb{R}^K \to \mathbb{R}^1$ are termed as *nodal* and *pool* functions, respectively, and assigned to the $k^{th}$ neuron of $l^{th}$ layer. In a heterogenous ONN configuration, every neuron has uniquely assigned $\psi$ and $P$ operators. Owing to this, an ONN network enjoys the flexibility of incorporating any non-linear transformation, which is suitable for the given learning problem. However, hand-crafting a suitable library of possible operators and searching for an optimal one for each neuron in a network introduces a significant overhead, which rises exponentially with increasing network complexity. Moreover, it is also possible that the ideal operator for the given learning problem cannot be expressed in terms of well-known functions. To resolve this key limitation, a composite nodal function is required that is iteratively created and tuned during back-propagation. A straightforward choice for accomplishing this would be to use a weighted combination of all operators in the operator set library and learning the weights during training. However, such a formulation would be susceptible to instability issues because of different dynamic ranges of individual functions. Additionally, it would still rely on the manual selection of suitable functions to populate the operator set library. Therefore, to formulate a nodal transformation that does not require any pre-selection and manual assignment of operators, we make use of the MacLaurin function approximation.

To formulate a nodal transformation, $\psi$ which does not require a pre-selection and manual assignment of operators, we use the MacLaurin based function approximation near the origin (x=0) as,

$$\psi(x) = \sum_{n=0}^{\infty} \frac{\psi^{(n)}(0)}{n!} x^n \quad (5)$$

The $Q^{th}$ order truncated approximation, formally known as the MacLaurin polynomial, takes the form of the following finite summation:

$$\psi(x)^{(Q)} = \sum_{n=0}^{Q} \frac{\psi^{(n)}(0)}{n!} x^n \quad (6)$$

The above formulation can approximate any function $\psi(x)$ sufficiently well near 0. When the activation function bounds the neuron's input feature maps in the vicinity of 0 (e.g., *tanh*) the formulation of (6) can be exploited to form a composite nodal operator where the power coefficients, $\frac{\psi^{(n)}(0)}{n!}$ can be the learned parameters of the network during training. It was shown in [57] that the nodal operator of the $k^{th}$ generative neuron in the $l^{th}$ layer can take the following general form:

$$\widetilde{\psi_k^l}\left(w_{ik}^{l(Q)}(r), y_i^{l-1}(m+r)\right)$$
$$= \sum_{q=1}^{Q} w_{ik}^{l(Q)}(r,q)\left(y_i^{l-1}(m+r)\right)^q \quad (7)$$

In (7), $Q$ is a hyperparameter which controls the degree of the MacLaurin series approximation, and $w_{ik}^{l(Q)}$ is a learnable kernel of the network. A key difference in (7) as compared to the convolutional (3) and operational (4) model is that $\widetilde{\psi_k^l}$ is not fixed, rather a distinct operator over each individual output, $y_i^{l-1}$, and thus requires $Q$ times more parameters. Therefore, the $K \times 1$ kernel vector $w_{ik}^l$ has been replaced by a $K \times Q$ matrix $w_{ik}^{l(Q)} \in \mathbb{R}^{K \times Q}$ which is formed by replacing each element $w_{ik}^l(r)$ with a $Q$-dimensional vector $w_{ik}^{l(Q)}(r) = [w_{ik}^{l(Q)}(r,1), w_{ik}^{l(Q)}(r,2), \dots, w_{ik}^{l(Q)}(Q)]$. The input map of the generative neuron, $\widetilde{x}_{ik}^l$ can now be expressed as,

$$\widetilde{x_{ik}^l}(m) = P_k^l \left( \sum_{q=1}^{Q} w_{ik}^{l(Q)}(r,q)\left(y_i^{l-1}(m+r)\right)^q \right)_{r=0}^{K-1} \quad (8)$$

During training, as $w_{ik}^{l(Q)}$ is iteratively tuned by the back-propagation (BP), *customized* nodal transformation functions will be generated as a result of (8), which would be uniquely tailored for $ik^{th}$ connection. This enables enhanced flexibility which provides three key benefits. Firstly, the need for manually defining a list of suitable nodal operators and searching for the optimal operator for each neuron connection is naturally alleviated. Secondly, the heterogeneity is not limited to each neuron connection $ik$ but down to each kernel element as $\widetilde{\psi}_l^k\left(w_{ik}^{l(Q)}(r), y_i^{l-1}(m+r)\right)$ will be unique $\forall r \in [0,1,\dots,K-1]$. As illustrated in Figure 2, such diversity is not achievable even with the flexible operational neuron model of ONNs. Thirdly, in generative neurons, the heterogeneity is driven only by the values of the weights $w_{ik}^{l(Q)}$ and the core operations (multiplication, summation) are the same for all neurons in a layer, as shown in (8). Owing to this, unlike ONNs, the generative neurons inside a Self-ONN layer can be parallelized much more efficiently, which leads to a considerable reduction in computational complexity and time. Moreover, a special case of (8) can also be expressed in terms of the widely applicable convolutional model.

### B. Representation in terms of convolution

If the pooling operator $P_k^l$ is fixed to summation operator, $\widetilde{x}_{ik}^l$ is then defined as:

$$\widetilde{x_{ik}^l}(m) = \sum_{r=0}^{K-1}\sum_{q=1}^{Q} w_{ik}^{l(Q)}(r,q)\left(y_i^{l-1}(m+r)\right)^q \quad (9)$$

Exploiting the commutativity of the summation operations in (9), we can alternatively write:

$$\widetilde{x_{ik}^l}(m) = \sum_{q=1}^{Q}\sum_{r=0}^{K-1} w_{ik}^{l(Q)}(r,q-1)y_i^{l-1}(m+r)^q \quad (10)$$

Using (1) and (2), the formula in (10) can be further simplified as follows:

$$\widetilde{x_{ik}^l} = \sum_{q=1}^{Q} Conv1D\left(w_{ik}^{l(Q)}, \left(y_i^{l-1}\right)^q\right) \quad (11)$$

Hence, the formulation can be accomplished by applying Q 1D convolution operations. If $Q$ is set to 1, (11) entails the convolutional formulation of (3). Therefore, as CNN is a subset of ONN corresponding to a specific operator set, it is also a special case of Self-ONN with $Q = 1$ for all neurons.

### C. Vectorized Notation

Expressing explicit loops in terms of matrix and vector manipulations is a key idea behind vectorization, which is a major driving factor behind fast implementations of modern-day neural network implementations. In this section, we first introduce how the vectorized notation can be used to express the 1D convolution operation inside a neuron. Afterward, the same key principles will be exploited to express the generative neuron formulation of (9) as a single matrix-vector product.

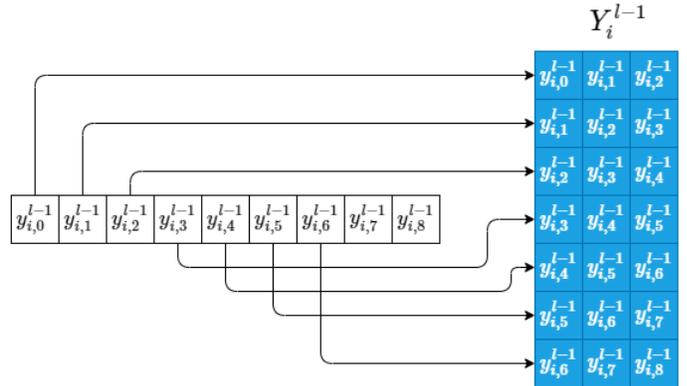

**Figure 3:** Reshuffling operation used to convert $y_i^{l-1}$ to $Y_i^{l-1}$.

First, an alternate formulation of the operation of (3) is now presented. We introduce a transformation $\delta(\cdot, K)$ which concatenates $y_i^{l-1}$ such that values inside each $K$-dimensional kernel as rows to form a matrix $Y_i^{l-1} \in \mathbb{R}^{M \times K}$. The process is visually depicted in Figure 3 for $K = 3$, and mathematically expressed in (12).

$$Y_i^{l-1} = \delta(y_i^{l-1}, K)$$
$$= \begin{bmatrix} y_i^{l-1}(0) & y_i^{l-1}(1) & \cdots & y_i^{l-1}(K-1) \\ \vdots & \vdots & \cdots & \vdots \\ y_i^{l-1}(m) & y_i^{l-1}(m+1) & \cdots & y_i^{l-1}(m+K-1) \\ \vdots & \vdots & \cdots & \vdots \\ y_i^{l-1}(M-1) & y_i^{l-1}(M) & \cdots & y_i^{l-1}(M+K-1) \end{bmatrix} \quad (12)$$

Secondly, we construct a matrix $W_{ik}^l \in \mathbb{R}^{M \times K}$ whose rows are repeated copies of $w_{ik} \in \mathbb{R}^K$.

$$W_{ik}^l = \begin{bmatrix} w_{ik}^l(0) & w_{ik}^l(1) & \cdots & w_{ik}^l(K-1) \\ \vdots & \vdots & \cdots & \vdots \\ w_{ik}^l(0) & w_{ik}^l(1) & \cdots & w_{ik}^l(K-1) \\ \vdots & \vdots & \cdots & \vdots \\ w_{ik}^l(0) & w_{ik}^l(1) & \cdots & w_{ik}^l(K-1) \end{bmatrix} \quad (13)$$

We now consider the Hadamard product of these two matrices:

$$\begin{aligned} Y_i^{l-1} &\otimes W_{ik}^l \\ &= \begin{bmatrix} y_i^{l-1}(0)w_{ik}^l(0) & \cdots & y_i^{l-1}(K-1)w_{ik}^l(K-1) \\ \vdots & \cdots & \vdots \\ y_i^{l-1}(m)w_{ik}^l(0) & \cdots & y_i^{l-1}(m+K-1)w_{ik}^l(K-1) \\ \vdots & \cdots & \vdots \\ y_i^{l-1}(M-1)w_{ik}^l(0) & \cdots & y_i^{l-1}(M+K-1)w_{ik}^l(K-1) \end{bmatrix} \end{aligned} \quad (14)$$

Applying the summation operation across rows, we get:

$$\sum \left( Y_i^{l-1} \otimes W_{ik}^l \right)(m) = \sum_{r=0}^{K-1} w_{ik}^l(r) y_i^{l-1}(m+r) \quad (15)$$

which is equivalent to (3). We also note that,

$$\sum Y_i^{l-1} \otimes W_{ik}^l = Y_i^{l-1} w_{ik}^l \quad (16)$$

Therefore,

$$x_{ik}^l(m) = \left( Y_i^{l-1} w_{ik}^l \right)(m) \quad (17)$$

$$x_{ik}^l = Y_i^{l-1} w_{ik}^l \quad (18)$$

Hence, the 1D convolution operation can be represented in terms of a single matrix-vector product. This operation lies at the heart of conventional explicit general matrix multiplications (GEMM) based convolution implementations and enables efficient usage of parallel computational resources such as GPU cores.

### D. Forward Propagation through a 1D Self-ONN neuron

We showed in (11) how the Self-ONN formulation of (10) can be represented as a summation of $Q$ individual convolutional operations. Moreover, from (12), a convolutional operation can be represented as a matrix-vector product. We now use these two formulations to represent the transformation of (11) as a single convolution operation, and consequently a single matrix-vector product, instead of Q-separate ones.

We start by introducing $Y_i^{l-1^{(Q)}} \in \mathbb{R}^{M \times KQ}$ such that

$$Y_i^{l-1^{(Q)}} = \begin{bmatrix} Y_i^{l-1} & \left(Y_i^{l-1}\right)^{\circ 2} & \cdots & \left(Y_i^{l-1}\right)^{\circ Q} \end{bmatrix} \quad (19)$$

where $\circ n$ is the Hadamard exponentiation operator. The $m^{th}$ row of $Y_i^{l-1^{(Q)}}$ can be expressed as,

$$Y_i^{l-1^{(Q)}}(m) = \begin{bmatrix} y_i^{l-1}(m) \\ \vdots \\ y_i^{l-1}(m+K-1) \\ \vdots \\ y_i^{l-1}(m)^2 \\ \vdots \\ y_i^{l-1}(m+K-1)^2 \\ \vdots \\ y_i^{l-1}(m)^Q \\ \vdots \\ y_i^{l-1}(m+K-1)^Q \end{bmatrix}^T \quad (20)$$

Moreover, we construct $W_{ik}^{l(Q)} \in \mathbb{R}^{M \times KQ}$ by first vectorizing $w_{ik}^{l(Q)} \in \mathbb{R}^{K \times Q}$ to $\overrightarrow{w_{ik}^{l(Q)}} \in \mathbb{R}^{KQ}$ and then concatenating $m$ copies of $\overrightarrow{w_{ik}^{l(Q)}}$ along the row dimension, as expressed in (21) and (22).

$$\overrightarrow{w_{ik}^{l(Q)}} = \begin{bmatrix} w_{ik}^{l(Q)}(0,0) \\ \vdots \\ w_{ik}^{l(Q)}(K-1,0) \\ w_{ik}^{l(Q)}(0,1) \\ \vdots \\ w_{ik}^{l(Q)}(K-1,1) \\ \vdots \\ w_{ik}^{l(Q)}(0,Q-1) \\ \vdots \\ w_{ik}^{l(Q)}(K-1,Q-1) \end{bmatrix}^T \quad (21)$$

$$W_{ik}^{l(Q)}(m) = \overrightarrow{w_{ik}^{l(Q)}} \quad (22)$$

Taking the Hadamard Product of $Y_i^{l-1^{(Q)}}$ and $W_{ik}^{l(Q)}$, we get:

$$\begin{aligned} &\left( Y_i^{l-1^{(Q)}} \otimes W_{ik}^{l(Q)} \right)(m) \\ &= \begin{bmatrix} y_i^{l-1}(m) w_{ik}^{l(Q)}(0,0) \\ \vdots \\ y_i^{l-1}(m+K-1) w_{ik}^{l(Q)}(K-1,0) \\ \vdots \\ y_i^{l-1}(m)^2 w_{ik}^{l(Q)}(0,1) \\ \vdots \\ y_i^{l-1}(m+K-1)^2 w_{ik}^{l(Q)}(K-1,1) \\ \vdots \\ y_i^{l-1}(m)^Q w_{ik}^{l(Q)}(0,Q-1) \\ \vdots \\ y_i^{l-1}(m+K-1)^Q w_{ik}^{l(Q)}(K-1,Q-1) \end{bmatrix}^T \end{aligned} \quad (23)$$

Summation of the above gives us

$$\left( \sum \left( Y_i^{l-1^{(Q)}} \otimes W_{ik}^{l(Q)} \right) \right)(m) \quad (24)$$

$$= \sum_{r=0}^{K-1} y_i^{l-1}(m+r) w_{ik}^{l(Q)}(r,0)$$
$$+ \sum_{r=0}^{K-1} y_i^{l-1}(m+r)^2 w_{ik}^{l(Q)}(r,0)$$
$$+ \ldots$$
$$+ \sum_{r=0}^{K-1} y_i^{l-1}(m+r)^Q w_{ik}^{l(Q)}(r, Q-1)$$
$$= \sum_{q=1}^{Q} \sum_{r=0}^{K-1} y_i^{l-1}(m+r)^q w_{ik}^{l(Q)}(r, q-1)$$

This is equivalent to (10). So, one can write,

$$\left( \sum \left( Y_i^{l-1(Q)} \otimes W_{ik}^{l(Q)} \right) \right)(m) = \widetilde{x_{ik}^l}(m) \quad (25)$$

Also, using (24), we can write,

$$\sum \left( Y_i^{l-1(Q)} \otimes W_{ik}^{l(Q)} \right)$$
$$= \begin{bmatrix} \sum_{q=1}^{Q} \sum_{r=0}^{K-1} y_i^{l-1}(r)^q w_{ik}^{l(Q)}(r, q-1) \\ \vdots \\ \sum_{q=1}^{Q} \sum_{r=0}^{K-1} y_i^{l-1}(m+r)^q w_{ik}^{l(Q)}(r, q-1) \\ \vdots \\ \sum_{q=1}^{Q} \sum_{r=0}^{K-1} y_i^{l-1}(M-1+r)^q w_{ik}^{l(Q)}(r, q-1) \end{bmatrix} \quad (26)$$

Finally, from (25) and (26), we can simply infer that:

$$\widetilde{x_{ik}^l} = Y_i^{l-1(Q)} \left( \overrightarrow{w_{ik}^{l(Q)}} \right) \quad (27)$$

The formulation of (27) provides a key computational benefit, as the forward propagation through the generative neuron is accomplished using a single matrix-vector multiplication. Hence, in theory, if the computational cost and memory requirement of constructing matrices $Y_i^{l-1(Q)}$ and $W_{ik}^{l(Q)}$ is considered negligible, the complexity of a convolutional neuron is approximately the same as that of the generative neuron, as both can be accomplished by a single matrix-vector product. Finally, to complete forward propagation, using (1), we can express:

$$\widetilde{x_k^l} = b_k^l + \sum_{i=0}^{N_{l-1}} \widetilde{x_{ik}^l} \quad (28)$$

### E. Back-propagation

We now proceed to derive the back-propagation formulation for the generative neuron model of 1D Self-ONN by utilizing the vectorized notation introduced in Section II.D. To back-propagate the error through the generative neuron, given the derivative of the loss w.r.t the neuron's output, $dL/d\widetilde{x_{ik}^l}$, we aim to define $dL/dy_i^{l-1}$, $dL/dw_{ik}^{l(Q)}$ and $dL/db_k^l$.

We start by taking the derivative of (27) w.r.t $\overline{Y_i^{l-1(Q)}}$ as follows:

$$\frac{d\widetilde{x_{ik}^l}(m)}{dY_i^{l-1(Q)}(\bar{m})} = \begin{cases} \overrightarrow{w_{ik}^{l(Q)}} & m = \bar{m} \\ \vec{0} & otherwise \end{cases} \quad (29)$$

Using (29), we can now apply the chain rule to get:

$$\frac{dL}{dY_i^{l-1(Q)}} = \frac{dL}{d\widetilde{x_{ik}^l}} \frac{d\widetilde{x_{ik}^l}}{dY_i^{l-1(Q)}} \quad (30)$$

Given $\frac{dL}{dY_i^{l-1(Q)}}$, we aim to find the derivative of loss w.r.t to the previous layer's output:

$$\frac{dL}{dy_i^{l-1}} = \frac{dL}{dY_i^{l-1}} \frac{dY_i^{l-1}}{dy_i^{l-1}} \quad (31)$$

We know from (19) that $Y_i^{l-1(Q)} = \begin{bmatrix} Y_i^{l-1} & (Y_i^{l-1})^{\circ 2} & \cdots & (Y_i^{l-1})^{\circ Q} \end{bmatrix}$. Taking the derivative of (19) w.r.t $Y_i^{l-1}$:

$$\frac{dY_i^{l-1(Q)}}{dY_i^{l-1}} = \begin{bmatrix} 1 & 2(Y_i^{l-1})^{\circ 1} & \cdots & Q(Y_i^{l-1})^{\circ Q-1} \end{bmatrix} \quad (32)$$

Using this we can write:

$$\frac{dL}{dY_i^{l-1}} = \frac{dL}{dY_i^{l-1(Q)}} \frac{dY_i^{l-1(Q)}}{dY_i^{l-1}} \quad (33)$$

Finally, we are able now to calculate the derivative of loss w.r.t $y_i^{l-1}$ as follows:

$$\frac{dL}{dy_i^{l-1}(\bar{m})} = \sum_{m=0}^{M-1} \frac{dL}{dY_i^{l-1}(m)} \frac{dY_i^{l-1}(m)}{dy_i^{l-1}(\bar{m})}$$
$$= \sum_{m=0}^{M-1} \frac{dL}{dY_i^{l-1}(m)} \left[ \frac{dy_i^{l-1}(m)}{dy_i^{l-1}(\bar{m})}, \ldots, \frac{dy_i^{l-1}(m+K-1)}{dy_i^{l-1}(\bar{m})} \right] \quad (34)$$

From (34) and (12), we can notice that $\left( \frac{dY_i^{l-1}(m)}{dy_i^{l-1}(\bar{m})} \right)$ will be equal to 1 only when the condition $m \leq \bar{m} \leq (m + K - 1)$ is met, and 0, otherwise. Moreover, as there are no repeating entries in each row of $Y_i^{l-1}$, only one element of $\frac{dY_i^{l-1}(m)}{dy_i^{l-1}(\bar{m})}$ can be non-zero and the location of this non-zero element is given by

$mod(\bar{m}, K)$. Based on these two points, we can infer the following:

$$\frac{dL}{dy_i^{l-1}(\bar{m})} = \sum_{m=0}^{M-1} \begin{cases} \frac{dL}{dY_i^{l-1}}(m, mod(\bar{m},K)) & m \leq \bar{m} \leq (m+K) \\ 0, & otherwise \end{cases} \quad (35)$$

The only other partial derivative needed for completing the back-propagation is the of the loss w.r.t the weights of the neuron $\overrightarrow{w_{\iota k}^{l(Q)}}$. By the chain rule, we can write:

$$\frac{dL}{dw_{\iota k}^{l(Q)}}(\bar{r}) = \frac{dL}{d\widetilde{x_{\iota k}^l}} \frac{d\widetilde{x_{\iota k}^l}}{dw_{\iota k}^{l(Q)}} \quad (36)$$

where $\frac{d\widetilde{x_{\iota k}^l}}{dw_{\iota k}^{l(Q)}}$ can be calculated by taking the derivative of (27) w.r.t $\overrightarrow{w_{\iota k}^{l(Q)}}$ as follows:

$$\frac{d\widetilde{x_{\iota k}^l}}{dw_{\iota k}^{l(Q)}} = Y_i^{l-1^{(Q)}} \quad (37)$$

For bias, we can use (28) to write:

$$\frac{dL}{db_k^l} = \frac{dL}{d\widetilde{x_k^l}} \frac{d\widetilde{x_k^l}}{db_k^l} = \sum_{m=0}^{M-1} \frac{dL}{d\widetilde{x_k^l}}(m) \quad (38)$$

Finally, assuming an SGD-based optimization, the weights and biases can be updated as follows:

$$\overrightarrow{w_{\iota k}^{l(Q)}}(t+1) = \overrightarrow{w_{\iota k}^{l(Q)}}(t) - \epsilon(t) \frac{dL}{dw_{\iota k}^{l(Q)}} \quad (39)$$

$$b_k^l(t+1) = b_k^l(t) - \epsilon(t) \frac{dL}{db_k^l} \quad (40)$$

where $\epsilon(t)$ is the learning factor at iteration $t$.

F. Computational Complexity

In this section, we provide the formulation for calculating the total number of multiply-accumulate operations (MACs) and the total number of parameters (PARs) of a generative neuron inside a 1D Self-ONN. To calculate the number of trainable parameters, we recall that, for each kernel connection, the generative neuron has $Q$ times more learnable parameters. Cumulatively, the number of trainable parameters, $n_k^l$, of the $k^{th}$ neuron of $l^{th}$ layer is given by the following formulation:

$$n_k^l = N_{l-1} * K_k^l * Q_k^l \quad (41)$$

In (41), $N_{l-1}$ is the number of neurons in layer $l-1$, $K_k^l$ is the kernel size used in the neuron and $Q_k^l$ is the approximation order selected for this neuron. Finally, to calculate the total number of MAC operations, one can note from (26) that to produce a single element in the output $\widetilde{x_{\iota k}^l}$, we require $K_k^l * Q_k^l$ MAC operations for each output map $y_i^{l-1}$ of the previous layer. Generalizing this, we can write the following:

$$MAC_k^l = N_{l-1} * \left| \widetilde{x_{\iota k}^l} \right| * K_k^l * Q_k^l \quad (42)$$

where $|\cdot|$ is the cardinality operator. For notational convenience, the bias term and the cost of Hadamard exponentiation are omitted from (42).

**Table I: Network models and their computational complexities. The average time corresponds to the time for classifying a single beat.**

| Network | Layer 1 | Layer 2 | PARs | MACs (M) | Avg. BP Time (µs) | Avg. FP Time (µs) |
|---|---|---|---|---|---|---|
| 1D CNN [27] | 32 | 16 | 8913 | 1451 | 54 | 8.3 |
| 1D Self-ONN (Q=7) | 16 | 8 | 16969 | 1843 | 114 | 21.2 |

We implemented the proposed patient-specific 1D ONN classifier using FastONN [48] and PyTorch machine learning libraries. All the experiments reported in this paper were run on a 2.2GHz Intel Core i7-8750H with 8 GB of RAM and NVIDIA GeForce GTX 1050Ti graphic card. Both training and testing phases of the classifier were processed by CUDA kernels. For calculation of inference times, an Intel Core i7-10750H CPU was used. Along with the average time complexity, using the formulations in (41) and (42), we provide the overall *PARs* and *MACs* for both network models in Table I.

Another important advantage of the proposed system is its significantly low computational cost for the ECG beat classification. Specifically, for a single CPU implementation, the total time for a single beat to BP per epoch and to FP to obtain the class vector is about 54 and 21.2 microseconds for a Q=7 Self-ONN classifier, respectively. Such an FP time to process a beat is indeed an insignificant complexity and naturally allows real-time implementation even on low-power mobile devices.

G. MIT/BIH Arrhythmia Dataset

The MIT/BIH arrhythmia dataset [59] is primarily used for performance evaluation in this study since it is the benchmark dataset that was used by all recent patient-specific ECG classification methods proposed in the literature. MIT/BIH dataset contains 48 ECG recordings, each with 30-min duration and two-channel ECG signals. Each record is extracted from the 24-hour ECG of 47 individuals. Each ECG record is pre-processed by band-pass filtering at 0.1-100 Hz and then sampled at 360 Hz. The database contains annotation for both timing information and beat class information verified by independent experts. To comply with the AAMI ECAR-1987 recommended practice [23], in this work, we apply the same data partitioning as in [26]-[29]. Accordingly, excluding the four which contain paced heartbeats, the remaining 44 ECG records from the MIT/BIH arrhythmia database are used for performance evaluation. Those records with patient IDs in the range of 100 to 124 encapsulating the common ECG patterns of routine clinical recordings are used to select the "common" ECG beats for training. Other records with patient IDs from 200 to 234 include less common but clinically significant arrhythmias such as ventricular, junctional, and supraventricular arrhythmias [59]. Overall, 83,648 ECG beats in those 44 recordings are used for testing and performance evaluation. According to the AAMI recommendation, we adopt the following five ECG beat categories: normal (N) beats originating in the sinus mode, supraventricular ectopic (S), ventricular ectopic (V), fusion (F), and unclassifiable (Q) beats. The beat detection method is outside the scope of this paper and there are numerous accurate beat detection algorithms proposed in the literature [58], [60].

**Table 2: Mapping the MIT-BIH Arrhythmia Database Heartbeat Types to the AAMI Heartbeat Classes [23].**

| AAMI Beat Class | MIT-BIH Normal/Arrhythmia types |
|---|---|
| **Non-ectopic beats (N)** | Normal (N) |
| | Right bundle branch block beats (RBBB) |
| | Atrial escape beats (e) |
| | Left bundle branch block (LBBB) |
| | Nodal (junctional) escape beats (j) |
| **Supraventricular ectopic beats (S)** | Aberrated atrial premature beats (a) |
| | Atrial premature contraction (A) |
| | Supraventricular premature beats (S) |
| | Nodal (junctional) premature beats (J) |
| **Ventricular ectopic beats (V)** | Premature ventricular contraction (PVC) |
| | Ventricular escape beats (E) |
| | Ventricular |
| | flutter wave (!) |
| **Fusion beats (F)** | Fusion of ventricular and normal beat (F) |
| **Unknown beats (Q)** | Paced beats (/) |
| | Fusion of paced and normal beats (f) |
| | Unclassifiable beats (Q) |

The raw data of each beat is represented by 128 samples via down-sampling which is the same resolution considered for the evaluation of the earlier 1D CNN counterpart [27] and previous works [26],[28],[29]. We also apply the same formation as in [27] for input ECG signals: to capture the morphological pattern, each beat is individually fed as the first channel of the network's input layer. To capture the temporal characteristics of each beat, a beat-trio is created from the neighbor beats and fed as the second channel of the input layer.

The training data for each patient is partitioned into two parts: global (the common set of 245 beats randomly selected from the ECG records of the patients with IDs 100-124) and patient-specific that belongs to the first 5 minutes of each patient's ECG record, which is the recommended practice by AAMI for training [23]. In brief, for each patient's ECG record there are usually around 500-700 beats used for training, and the rest (after the first 5 minutes) of the ECG record is used for the test.

## III. RESULTS

This section will first outline the benchmark ECG dataset used in this study and present the experimental setup used for testing and evaluation of the proposed patient-specific ECG classification based on 1D Self-ONNs. An extensive set of ECG classification experiments and comparative evaluations against the recent methods over the benchmark MIT-BIH dataset will be presented next. Finally, the computational complexity analysis of the 1D Self-ONNs and 1D CNNs will be formulated in detail.

### A. Experimental Setup

We aim to have a fair comparison with the earlier 1D CNN counterpart [27]. Therefore, in all experiments, we used a compact 1D Self-ONN consisting of only two Self-ONN layers and two dense (MLP) layers. However, we halved the number of neurons used in [27], i.e., each 1D Self-ONN used has 16 and 8 neurons on the first and second hidden layers, respectively, and 10 neurons on the hidden MLP layer. The output MLP layer size is 5 which is the number of ECG classes and the input layer size is two for single and beat-trio raw data representation. The same kernel sizes and the sub-sampling factors which are 15 and 6, respectively in [27] are used. As a result, the sub-sampling factor for the last Self-ONN layer is set to 5. Once detected, each ECG beat (single or beat-trio) is resized to 128 samples.

For all experiments, as in [27] a shallow training is employed by using Stochastic Gradient Descent (SGD), and a two-fold stopping criterion is used: the maximum number of BP epochs is 50 or minimum (train) classification error level is 3%. Therefore, training terminates as soon as either of the criteria is met. We initially set the learning rate, $\epsilon(0)$, as 0.01 and it is globally adapted during each BP iteration. We performed five individual BP runs for training over each patient's data and for comparative evaluations, report the best classification performance.

### B. ECG Classification Performance Evaluation

The experiments are performed over 44 patient records in the MIT/BIH arrhythmia database, which contains a total of 100,389 ECG beats. We intentionally employed the same patient-specific training methodology applied in [27]. For training the 1D Self-ONNs, both global and patient-specific training patterns are used. The common part of the training data set contains a total of 245 representative beats, including 75 from each type N, -S, and -V beats, and all 13 type-F and 7 type-Q beats, *randomly* selected from the records with patient IDs from 100 to 124 in MIT/BIH dataset.

In this study, the following standard metrics are used: classification accuracy ($Acc$), sensitivity ($Sen$), specificity ($Spe$), positive predictivity ($Ppr$), and F1-score ($F1$). Since there is a large variation in the number of beats from different classes (class imbalance) in the training/testing data (i.e. 39465/50354 type-N, 1277/5716 type-V, and 190/2571 type-S beats), sensitivity, specificity, positive predictivity and especially, F1-score are all relevant performance criteria for medical diagnosis applications.

For performance evaluation of the 1D Self-ONN classifier, we shall consider the two-channel (the single beat and the beat trio) raw ECG data segments of 128 samples long as the input. For all 44 records in the MIT/BIH arrhythmia database, the average ECG beat classification performance of the *state-of-the-art* 1D CNN method in [26] and 1D Self-ONN classifier for different Q values is presented in Table III. In SVEB classification, Self-ONNs have a significant superiority in all metrics yielding around 7% F1 gain over the CNN in [27]. In VEB classification, the gap in F1 still exists but is lower (1.5%). In order to perform a more extensive comparative evaluation, the performance metrics of the proposed approach is compared with the four existing algorithms, [13], [25]-[27], and [31] all of which use the same training and test datasets and comply with the AAMI standards. However, [21], [61]-[63] do not obey AAMI standards for patient-specific train set partitioning. Moreover, many of the recent studies use a large common train data (e.g., [33]-[37], [61]-[63]) for training and in fact, some even use data from 200 ID patients which should entirely be left for testing.

Following the AAMI recommendations, the problem of VEB and SVEB detection is considered individually and the results are summarized in Table IV. As the table details, MIT/BIH dataset is partitioned into three evaluation datasets: Dataset 1 contains 14 test recordings 11 of which (200, 202, 210, 213, 214, 219, 221, 228, 231, 233, and 234) are used for VEB detection while for SVEB detection, the remaining records 212, 222, and 232, are also used. Dataset 1 is common for all competing methods. Dataset 2 contains 24 records with patient IDs 200 and higher. Four competing methods (the proposed, [25], [26], [27], and [31]) are tested on this dataset. The recent method [31] indeed follows the AAMI standard and performed the same data partitioning for train and test sets of each patient. However, this study performed a dedicated search to select the most suitable common training data that can represent the beats of each patient, i.e., [31] performed an iterative search (repeated 200 times) for this purpose in order to select the best set of S beats whereas all other methods including the proposed approach randomly selected the common beats. Despite this advantage which makes a direct comparison unfair, the proposed method still achieves a similar or better performance than [31]. Finally, Dataset 3 is the entire MIT-BIH dataset with 44 records over which three methods, the proposed, [26] and [27], are compared. Finally, we present the distribution of beats and classification results per patient in Appendix B.

Table III: Average ECG beat classification performance of Kiranyaz *et al.* [27], and 1D Self-ONN classifiers for different $Q$ values for all 44 records in the MIT/BIH arrhythmia database.

| Q | SVEB | | | VEB | | |
|---|---|---|---|---|---|---|
| | *Sen* | *Ppr* | *F1* | *Sen* | *Ppr* | *F1* |
| [27] | 0.603 | 0.635 | 0.618 | **0.939** | 0.906 | 0.922 |
| 3 | 0.620 | **0.767** | **0.686** | 0.893 | 0.951 | 0.921 |
| 5 | 0.626 | 0.634 | 0.630 | 0.900 | 0.956 | 0.927 |
| 7 | 0.617 | 0.749 | 0.677 | 0.918 | **0.957** | **0.937** |
| 9 | **0.634** | 0.717 | 0.673 | 0.885 | 0.947 | 0.915 |

Table IV: VEB and SVEB classification performances of the proposed 1D Self-ONN with $Q = 7$ and five competing algorithms. Best results are highlighted.

| Methods | SVEB | | | | | VEB | | | | |
|---|---|---|---|---|---|---|---|---|---|---|
| | *Acc* | *Sen* | *Spe* | *Ppr* | *F1* | *Acc* | *Sen* | *Spe* | *Ppr* | *F1* |
| Hu et al [13][1] | N/A | N/A | N/A | N/A | N/A | 94.8 | 78.9 | 96.8 | 75.8 | 77.3 |
| Jiang and Kong [25][1] | 97.5 | 74.9 | 98.8 | 78.8 | 76.8 | 98.8 | 94.3 | 99.4 | 95.8 | 95.0 |
| Ince et al. [26][1] | 96.1 | 81.8 | 98.5 | 63.4 | 71.4 | 97.9 | 90.3 | 98.8 | 92.2 | 91.2 |
| Kiranyaz et al. [27][1] | 96.4 | 68.8 | 99.5 | 79.2 | 73.6 | 98.9 | 95.9 | 99.4 | 96.2 | 96.0 |
| Zhai et al. [31][1] | 97.3 | 85.3 | 98.0 | 71.8 | 77.9 | **99.1** | **96.4** | **99.5** | 96.4 | 96.4 |
| **Proposed method**[1] with $Q = 7$ | **97.3** | **76.0** | **99.6** | **82.2** | **78.9** | 99.0 | 95.3 | 98.9 | **97.9** | **96.5** |
| Jiang and Kong [25][2] | 96.6 | 50.6 | 98.8 | 67.9 | 57.9 | 98.1 | 86.6 | 99.2 | 93.3 | 89.8 |
| Ince et al. [26][2] | 96.1 | 62.1 | 98.5 | 56.7 | 59.2 | 97.5 | 83.4 | 98.1 | 87.4 | 85.3 |
| Kiranyaz et al. [27][2] | 96.4 | 64.6 | 98.6 | 62.1 | 63.3 | 98.6 | 95.0 | 98.1 | 89.5 | 92.1 |
| Zhai et al. [31][2] | **97.5** | **76.8** | 98.7 | 74.0 | **75.3** | 98.6 | **93.8** | 99.2 | 92.4 | 93.0 |
| **Proposed method**[2] with $Q = 7$ | 97.0 | 60.6 | **99.5** | **79.3** | 68.7 | **98.7** | 91.0 | **99.2** | **96.6** | **93.7** |
| Ince et al. [26][3] | 97.4 | **63.5** | 99.0 | 53.7 | 58.1 | 98.3 | 84.6 | 98.7 | 87.4 | 85.9 |
| Kiranyaz et al. [27][3] | 97.6 | 60.3 | 99.2 | 63.5 | 61.8 | 99.0 | **93.9** | 98.9 | 90.6 | 92.2 |
| **Proposed method**[3] with $Q = 7$ | **98.0** | 61.7 | **99.6** | **74.8** | **67.6** | **99.04** | 91.8 | **99.4** | **95.7** | **93.7** |

[1]The comparison results are based on 11 common recordings for VEB detection and 14 common recordings for SVEB detection.
[2]The VEB and SVEB detection results are compared for 24 common testing records only.
[3]The VEB and SVEB detection results over all records in the MIT-BIH dataset.

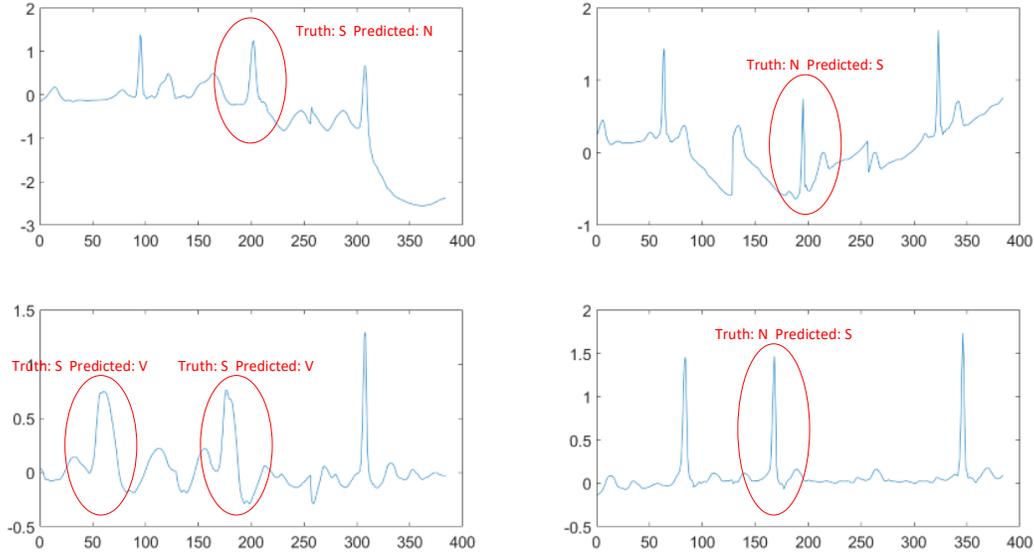

**Figure 4: Four ECG samples from the test section of the patient 202's ECG record with the ground-truth labels.**

## IV. DISCUSSION

The results presented in Table IV show that for all three dataset partitions, sensitivity (recall) and positive predictivity (precision) rates of SVEB detection are significantly lower than VEB detection. The main reason is that SVEB class is under-represented in the training data and hence more S beats are misclassified as N or V beats. The proposed 1D Self-ONN classifier significantly improves both precision and F1-score of SVEB and VEB detections for all data partitions compared to [27] and other older competing algorithms. It is also observed that, for SVEB detection over the entire dataset, while the *sensitivity* of the proposed method is slightly lower than [26], *precision* is significantly higher (> 20%). This means that false positives are significantly reduced by 1D Self-ONNs. For the VEB classification over the entire dataset, the *sensitivity* of the proposed method is slightly lower than [27], *precision* is significantly higher (> 5%). Therefore, the same conclusion can be drawn for [27] too.

**Table V: Confusion matrix for the ECG beat classification of Patient 202. The results from [27] are shown between parentheses.**

|  |  | Classification Result | | | | |
|---|---|---|---|---|---|---|
|  |  | N | S | V | F | Q |
| Ground Truth | N | **1493** (1435) | 203 (258) | 6 (10) | 0 (0) | 0 (0) |
|  | S | 20 (11) | **23** (15) | 12 (28) | 0 (0) | 0 (0) |
|  | V | 1 (4) | 3 (1) | **11** (**10**) | 0 (0) | 0 (0) |
|  | F | 1 (1) | 0 (0) | 0 (0) | 0 (0) | 0 (0) |
|  | Q | 0 (0) | 0 (0) | 0 (0) | 0 (0) | 0 (0) |

The superiority of the proposed approach over the landmark study, [27], is obvious. Among all competing methods, the one with the closest performance level with the proposed work is [31]. However, in [31] the authors performed a dedicated search to select the most suitable common training data that can represent the beats of each patient, e.g., the authors performed an iterative search (to be repeated 200 times) for the same purpose in order to select the best set of S beats. Although this makes it unfair to compare with the proposed and other methods where the common beats are *randomly* selected, we still decided to include this work in our comparative evaluations. Moreover, there are some challenging ECG records particularly on the test dataset, e.g., from the patients with IDs: 201, 202, 209, 222, and 232. For these ECG signals, both the patient-specific data from the first 5 minutes interval and the common data of 245 beats extracted randomly from the training dataset do not successfully characterize the patient's S and/or V beats. Take for instance the confusion matrix data resulting for test patient 202 in Table V using both 1D Self-ONN and 1D CNN classifiers. While in 1D CNN, 258 normal beats are misclassified as S beats and 28 S beats are misclassified as V beats, the proposed classifier misclassifies 203 normal beats as S beats and 12 S beats as V beats. On the other hand, the corresponding trade-off is that 20 S beats are misclassified as normal compared to 11 for the 1D CNN classifier. The four ECG intervals from the test section of this patient's ECG record are shown in Figure 4. The characteristics of the ECG beats shown in the plots are quite different from the ones in the training dataset of this patient and hence such misclassifications occurred. For instance, the anomalies shown in the plot in the bottom-left were misclassified as a V beat due to its morphological variation, and the S beat in the bottom-right was misclassified as an N beat due to its temporal variation, which is a clear indication of an S beat.

Overall, the novel and significant contributions of this work can be enlisted as follows:

1. This is the first study that proposes 1D Self-ONNs over a 1D signal (ECG) repository. This is also the first study that proposes 1D Self-ONNs for a biomedical application.
2. We derive 1D vectorized formulations of Self-ONNs for forward- and back-propagation (BP) which lead to an efficient parallelization on dedicated hardware resources such as GPUs.
3. This is the first study that proposes and evaluates Self-ONNs for a classification task where a dense layer is used after the operational layers.
4. Finally, the proposed approach with the proposed 1D Self-ONN classifier significantly outperformed the *state-of-the-art* approach in the landmark study in [27] without a significant increase in the computational complexity.

The optimized PyTorch implementation of 1D Self-ONNs is publicly shared in 0.

## V. CONCLUSIONS

In this work, we proposed 1D Self-ONNs for patient-specific ECG classification and arrhythmia detection. Our objective is to push the frontier set by the landmark study [27] based on 1D CNNs by achieving a new *state-of-the-art* ECG classification performance with a superior computational efficiency. A Self-ONN is basically composed of generative neurons each of which is capable of optimizing the nodal operator function of each kernel operating over the individual output map of the previous layer neuron. This is indeed a neuron-level heterogeneity that maximizes the network diversity and thus the learning performance. Besides, Self-ONNs present other important advantages over conventional ONNs. First and foremost, its "self-organizing" capability voids the need for prior operator search. Moreover, in a Self-ONN, the nodal operator of each neuron of each output layer neuron, i.e., the most crucial neuron in the network in which the loss (fitness) is computed, can be optimized. Finally, due to the "on-the-fly" generation of the non-linear nodal operator, the network can create the best possible basis functions such that when operated with the normal and arrhythmic ECG signal, one can get the highest discrimination. Overall, in this new-generation machine learning paradigm, traditional weight optimization of conventional CNNs is entirely turned into an operator optimization process. Nevertheless, as demonstrated in this study, Self-ONNs can still be implemented by only 1D convolutions and in a parallelized implementation, a Self-ONN and a CNN with the same configuration have similar computational complexity.

An extensive set of comparative evaluations over the benchmark MIT/BIH arrhythmia database shows that for all test partitions, 1D Self-ONNs surpass 1D CNNs in both VEB and SVEB detection even with less number of learning units (half the number of neurons used in [27]). In particular, thanks to its improved learning performance, a significant improvement in precision and F1-score metrics of SVEB detection is achieved. This clearly shows that especially for low-power, mobile platforms, such as Holter devices, and when labeled data is scarce, the proposed approach can conveniently be used as a patient-specific ECG monitoring in real-time.

## APPENDIX A

### OPERATIONAL NEURAL NETWORKS

ONNs are derived from the Generalized Operational Perceptrons (GOPs) in the same way CNNs are derived from MLPs with two restrictions: limited connectivity and weight sharing. GOPs have been proposed in [40], [41] to replace the basic (linear) neuron model from the 1950s (McCulloch-Pitts) [39] aiming to address the well-known limitations and drawbacks of MLPs. Recently, GOPs have outperformed not only MLPs but even the latest variants of Extreme Learning Machines (ELMs) in [42]-[45]. Derived directly from GOPs, ONNs [46]-[49] are heterogeneous networks encapsulating neurons with linear and non-linear operators hence carrying a closer link to biological systems. In brief, ONNs extend the sole usage of linear convolutions in the convolutional neurons by the *nodal* and *pool* operators. Conventional 2D CNNs and ONNs are illustrated in Figure 5 where the right side illustrates a three operational layer ONN with a 3x3 kernel in each neuron. As illustrated in the figure, the input map of the $k^{th}$ neuron at the current layer, $x_k^l$, is obtained by *pooling* the final output maps, $y_i^{l-1}$ of the previous layer neurons *operated* with its corresponding kernels, $w_{ki}^l$, as follows:

$$x_k^l = b_k^l + \sum_{i=1}^{N_{l-1}} oper2D(w_{ki}^l, y_i^{l-1}, 'NoZeroPad')$$

$$x_k^l(m,n)\Big|_{(0,0)}^{(M-1,N-1)} = b_k^l + \qquad (43)$$

$$\sum_{i=1}^{N_{l-1}} \left( P_k^l \begin{bmatrix} \Psi_{ki}^l\left(w_{ki}^l(0,0), y_i^{l-1}(m,n)\right), \dots, \\ \Psi_{ki}^l(w_{ki}^l(r,t), y_i^{l-1}(m+r,n+t)), \dots \end{bmatrix} \right)$$

A close look at Eq. (43) reveals the fact that when the pool operator is "summation", $P_k^l = \Sigma$, and the nodal operator is "linear", $\Psi_{ki}^l(y_i^{l-1}(m,n), w_{ki}^l(r,t)) = w_{ki}^l(r,t)y_i^{l-1}(m,n)$, for *all* neurons, then the resulting homogenous ONN will be identical to a CNN (illustrated on the left side of the figure). Hence, ONNs are indeed a superset of CNNs as the GOPs are a superset of MLPs. The ultimate question one would ask over a heterogeneous network is how to determine the synaptic connections of each neuron, that is the assignment of the "right" operator set(s) for each neuron/layer. This is not a problem for a CNN simply because all neurons are linear, and one can only learn to alter the kernel weights. In [46], Greedy Iterative Search (GIS) has been proposed to find the best operator set per layer; however, GIS is computationally demanding and as a local search method, it may fail to find the global optimal sets. Moreover, ONNs with GIS-assigned operator sets will still be homogenous, layer-wise, since the same operator set is used for all neurons in each layer and its heterogeneity is further limited due to the finite number of operators in the operator set library. The latter can particularly cause performance degradation if the right operator set for the problem at hand is missing from the library.

Convolutional Layers of CNNs | Operational Layers of ONNs

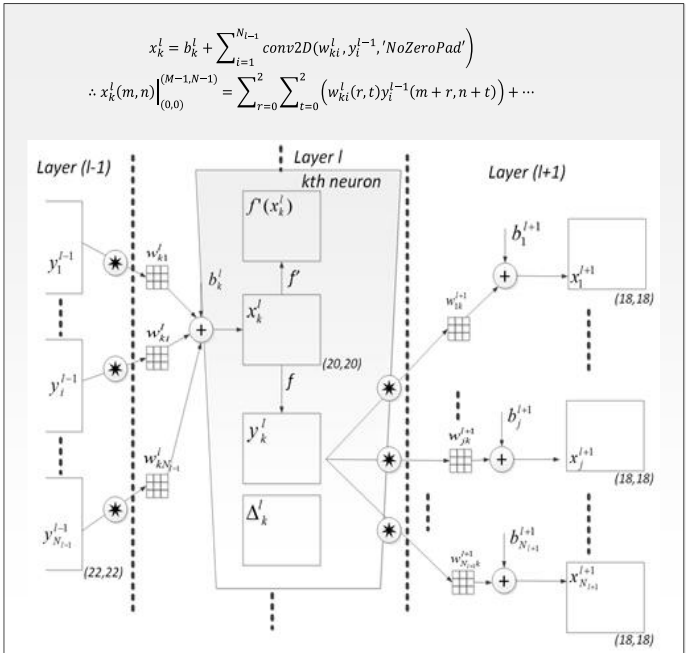
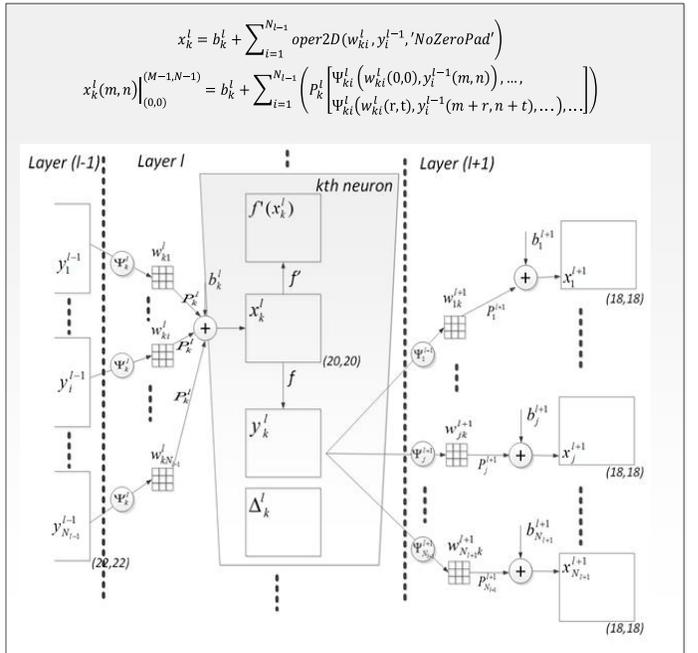

**Figure 5:** The illustration of the $k^{th}$ neuron of a CNN (left) and an ONN (right) along with the three consecutive convolutional (left) and operational (right) layers [46].

APPENDIX B

In this appendix, we present the distribution of beats and classification results per patient in Table 6.

Table 6: The distribution of beats per class and classification results per patient

| Patient Number | Number of Beats | | | | | | SVEB | | | | VEB | | | | |
|---|---|---|---|---|---|---|---|---|---|---|---|---|---|---|---|
| | N | S | V | F | Q | Acc | Sen | Spe | Ppr | F1 | Acc | Sen | Spe | Ppr | F1 |
| 100 | 1858 | 29 | 1 | 0 | 0 | 99.6 | 72.4 | 100.0 | 100.0 | 83.9 | 99.9 | 0.0 | 100.0 | - | - |
| 101 | 1545 | 3 | 0 | 0 | 2 | 99.8 | 33.3 | 99.8 | 50.0 | 39.9 | 99.9 | - | 99.9 | 0.0 | - |
| 103 | 1730 | 2 | 0 | 0 | 0 | 99.7 | 0.0 | 100.0 | 0.0 | - | 100.0 | - | 99.8 | - | - |
| 105 | 2104 | 0 | 28 | 0 | 5 | 98.4 | - | 97.6 | 0.0 | - | 97.0 | 85.7 | 99.1 | 33.3 | 47.9 |
| 106 | 1226 | 0 | 457 | 0 | 0 | 100.0 | - | 99.9 | 0.0 | - | 99.5 | 97.8 | 100.0 | 99.8 | 98.7 |
| 108 | 1446 | 3 | 13 | 2 | 0 | 99.0 | 33.3 | 99.0 | 7.7 | 12.5 | 98.4 | 46.2 | 99.2 | 28.6 | 35.3 |
| 109 | 2101 | 0 | 32 | 0 | 0 | 100.0 | - | 100.0 | - | - | 99.8 | 93.8 | 100.0 | 96.8 | 95.2 |
| 111 | 1764 | 0 | 1 | 0 | 0 | 99.9 | - | 99.9 | 0.0 | - | 99.9 | 100.0 | 99.9 | 50.0 | 66.6 |
| 112 | 2107 | 2 | 0 | 0 | 0 | 100.0 | 50.0 | 100.0 | 100.0 | 66.6 | 100.0 | - | 100.0 | - | - |
| 113 | 1486 | 5 | 0 | 0 | 0 | 99.7 | 100.0 | 100.0 | 50.0 | 66.6 | 100.0 | - | 99.7 | - | - |
| 114 | 1516 | 11 | 29 | 4 | 0 | 99.3 | 27.3 | 99.8 | 50.0 | 35.3 | 99.7 | 93.1 | 99.9 | 93.1 | 93.1 |
| 115 | 1622 | 0 | 0 | 0 | 0 | 99.8 | - | 100.0 | 0.0 | - | 99.9 | - | 99.8 | - | - |
| 116 | 1903 | 1 | 98 | 0 | 0 | 99.6 | 100.0 | 99.2 | 20.0 | 33.3 | 98.9 | 96.9 | 99.8 | 85.6 | 90.9 |
| 117 | 1273 | 1 | 0 | 0 | 0 | 99.9 | 0.0 | 99.8 | 0.0 | - | 99.9 | - | 99.9 | 0.0 | - |
| 118 | 1798 | 82 | 13 | 0 | 0 | 99.7 | 98.8 | 100.0 | 84.4 | 91.0 | 100.0 | 15.4 | 99.8 | 100.0 | 26.6 |
| 119 | 1288 | 0 | 363 | 0 | 0 | 100.0 | - | 100.0 | - | - | 99.8 | 99.2 | 100.0 | 100.0 | 99.5 |
| 121 | 1545 | 1 | 1 | 0 | 0 | 99.8 | 100.0 | 99.7 | 25.0 | 40.0 | 99.7 | 100.0 | 99.8 | 20.0 | 33.3 |
| 122 | 2057 | 0 | 0 | 0 | 0 | 100.0 | - | 100.0 | - | - | 100.0 | - | 100.0 | - | - |
| 123 | 1258 | 0 | 3 | 0 | 0 | 100.0 | - | 100.0 | - | - | 99.8 | 0.0 | 100.0 | - | - |
| 124 | 1291 | 8 | 40 | 4 | 0 | 94.0 | 12.5 | 99.8 | 1.4 | 2.5 | 99.4 | 85.0 | 94.7 | 97.1 | 90.6 |
| 200 | 1431 | 28 | 699 | 2 | 0 | 99.2 | 7.1 | 98.5 | 33.3 | 11.7 | 98.9 | 97.1 | 99.9 | 97.0 | 97.0 |
| 201 | 1308 | 125 | 198 | 2 | 0 | 92.1 | 2.4 | 98.9 | 18.8 | 4.2 | 96.5 | 69.2 | 99.8 | 90.7 | 78.5 |
| 202 | 1702 | 55 | 14 | 1 | 0 | 87.1 | 41.8 | 98.8 | 10.0 | 16.1 | 99.5 | 71.4 | 88.0 | 35.7 | 47.6 |
| 203 | 2099 | 0 | 373 | 1 | 4 | 99.5 | - | 97.7 | 0.0 | - | 96.4 | 89.0 | 99.7 | 87.8 | 88.3 |
| 205 | 2123 | 3 | 70 | 11 | 0 | 99.3 | 0.0 | 100.0 | 0.0 | - | 99.1 | 67.1 | 100.0 | 100.0 | 80.3 |
| 207 | 1336 | 96 | 109 | 0 | 0 | 93.1 | 0.0 | 99.3 | 0.0 | - | 93.1 | 15.6 | 99.8 | 65.4 | 25.1 |
| 208 | 1318 | 2 | 829 | 303 | 2 | 96.2 | 0.0 | 98.4 | 0.0 | - | 97.4 | 97.8 | 99.5 | 99.1 | 98.4 |
| 209 | 2125 | 371 | 1 | 0 | 0 | 88.9 | 29.4 | 100.0 | 88.6 | 44.1 | 100.0 | 0.0 | 99.3 | - | - |
| 210 | 2009 | 20 | 163 | 9 | 0 | 98.9 | 20.0 | 99.3 | 40.0 | 26.6 | 98.7 | 84.0 | 99.9 | 93.8 | 88.6 |
| 212 | 2282 | 0 | 0 | 0 | 0 | 100.0 | - | 100.0 | - | - | 100.0 | - | 100.0 | - | - |
| 213 | 2207 | 26 | 197 | 271 | 0 | 95.7 | 0.0 | 99.5 | 0.0 | - | 96.1 | 83.2 | 100.0 | 100.0 | 90.8 |
| 214 | 1661 | 0 | 213 | 1 | 2 | 99.8 | - | 99.9 | - | - | 99.6 | 97.7 | 100.0 | 100.0 | 98.8 |
| 215 | 2661 | 1 | 131 | 1 | 0 | 99.9 | 0.0 | 100.0 | - | - | 99.9 | 98.5 | 100.0 | 100.0 | 99.2 |
| 219 | 1731 | 7 | 50 | 0 | 0 | 99.4 | 0.0 | 99.8 | 0.0 | - | 99.0 | 74.0 | 99.8 | 90.2 | 81.3 |
| 220 | 1607 | 93 | 0 | 0 | 0 | 97.8 | 62.4 | 99.9 | 95.1 | 75.3 | 100.0 | - | 99.8 | 0.0 | - |
| 221 | 1699 | 0 | 318 | 0 | 0 | 100.0 | - | 100.0 | 0.0 | - | 99.8 | 98.4 | 100.0 | 100.0 | 99.1 |
| 222 | 1853 | 209 | 0 | 0 | 0 | 89.5 | 0.5 | 99.2 | 11.1 | 0.9 | 99.2 | - | 99.6 | 0.0 | - |
| 223 | 1635 | 66 | 454 | 8 | 0 | 96.8 | 74.2 | 99.6 | 54.4 | 62.7 | 96.8 | 85.5 | 97.9 | 98.7 | 91.6 |
| 228 | 1401 | 3 | 302 | 0 | 0 | 99.8 | 33.3 | 100.0 | 50.0 | 39.9 | 99.8 | 98.7 | 99.9 | 100.0 | 99.3 |
| 230 | 1872 | 0 | 1 | 0 | 0 | 100.0 | - | 100.0 | - | - | 99.9 | 0.0 | 100.0 | - | - |
| 231 | 1304 | 0 | 0 | 0 | 0 | 100.0 | - | 100.0 | - | - | 100.0 | - | 100.0 | - | - |
| 232 | 317 | 1162 | 0 | 0 | 0 | 97.7 | 99.1 | 100.0 | 98.0 | 98.5 | 100.0 | - | 92.4 | - | - |
| 233 | 1855 | 4 | 692 | 6 | 0 | 99.7 | 100.0 | 99.9 | 57.1 | 72.6 | 99.5 | 98.4 | 100.0 | 100.0 | 99.1 |
| 234 | 2234 | 50 | 3 | 0 | 0 | 97.9 | 4.0 | 99.9 | 100.0 | 7.6 | 99.8 | 0.0 | 100.0 | 0.0 | - |